\documentclass[twocolumn,superscriptaddress,showpacs,prl]{revtex4}
\usepackage{epsfig}
\usepackage{bm}

\newcommand{\mol}[3]{Mol. Phys. \textbf{#1}, {#2} (#3)}
\newcommand{\appl}[3]{Appl. Phys. Lett. \textbf{#1}, {#2} (#3)}
\newcommand{\phrb}[3]{Phys. Rev. B \textbf{#1}, {#2} (#3)}
\newcommand{\phre}[3]{Phys. Rev. E \textbf{#1}, {#2} (#3)}
\newcommand{\phrl}[3]{Phys. Rev. Lett. \textbf{#1}, {#2} (#3)}
\newcommand{\nature}[3]{Nature \textbf{#1}, {#2} (#3)}
\newcommand{\jetp}[3]{J. Exp. Theor. Phys. \textbf{#1}, {#2} (#3)}
\newcommand{\jpsj}[3]{J. Phys. Soc. Japan \textbf{#1}, {#2} (#3)}
\newcommand{\jetpl}[3]{JETP Lett. \textbf{#1}, {#2} (#3)}
\newcommand{\jpcm}[3]{J. Phys. Condens. Matter \textbf{#1}, {#2} (#3)}
\newcommand{\be}{\begin{equation}}
\newcommand{\en}{\end{equation}}
\newcommand{\bega}{\begin{eqnarray}}
\newcommand{\eda}{\end{eqnarray}}

\begin{document}
\title{Optical Magnetoelectric Effect in Multiferroic Materials: \\
Evidence for a Lorentz Force Acting on a Ray of Light}

\author{Kei Sawada}
\email{sawada@appi.t.u-tokyo.ac.jp}
\affiliation{Department of Applied Physics, the University of Tokyo, 
7-3-1, Hongo, Bunkyo-ku, Tokyo 113-8656, Japan}

\author{Naoto Nagaosa}
\email{nagaosa@appi.t.u-tokyo.ac.jp} 
\affiliation{Department of Applied Physics, the University of Tokyo, 
7-3-1, Hongo, Bunkyo-ku, Tokyo 113-8656, Japan}
\affiliation{Correlated Electron Research Center (CERC), 
National Institute of Advanced Industrial Science and Technology (AIST), 
Tsukuba Central 4, Tsukuba 305-8562, Japan}
\affiliation{CREST, Japan Science and Technology Agency (JST), Japan}

\date{\today}

\begin{abstract}

We theoretically propose that the 
optical analog of a Lorentz force acting on a ray of light is 
realized in multiferroic materials such as GaFeO$_3$ showing 
the magnetoelectric effect. The toroidal moment 
${\vec T} = \sum_j {\vec r}_j \times {\vec S}_j$ plays the role
of a ``vector potential" 
while its rotation corresponds to a ``magnetic field" for photons.
Hence, the light is subject to the Lorentz force when propagating through the 
domain wall region of the ferromagnetic or ferroelectric order. 
A realistic estimate on the magnitude of this effect is given. 
\end{abstract}

\maketitle


The analogy between optics and particle dynamics has been a
guiding principle in physics even leading to wave mechanics, i.e.,
quantum mechanics. 
In the classical theory, their analogy 
is established through the variational principle, which gives 
the Newtonian equation of motion for a particle 
and the eikonal equation of light. 
The variational principle for a classical particle is given by 
$\delta \int v ds=0$ with the velocity
$v = \sqrt{2[E-U(x)]/m}$ ($E$: total energy, $U(x)$: potential 
energy, $m$: mass of the particle). 
In geometrical optics, an equation for the 
trajectory of light is derived from the Fermat's  
principle; i.e., the time required to 
propagate from one to the other points is minimized.
$ \delta \int ds /V = \delta \int ds n({\vec r})/c =0$,
where $V=c/n$ is the velocity of light in the 
medium of refractive index $n({\vec r)}$.
This leads to the equation for the trajectory 
of light, which is similar to the equation of 
motion for a particle, as  
\begin{equation}
\frac{d}{ds}\bigl[ n(\vec{r}) \frac{d\vec{r}}{ds} \bigr] 
={\rm grad}\ n(\vec{r}).
\end{equation} 
In analogy to particle mechanics, 
we can interpret the above equation for light as follows. 
The term in the left-hand side is an ``acceleration term". 
On the right-hand side, 
the gradient of the refractive index $\nabla n({\vec r})$
acts as the ``force" on the light and distorts the trajectory, 
giving rise to the refraction of light, i.e., Snell's law. 
Therefore $n({\vec r})$ is regarded as the ``scalar potential".
In particle mechanics, there is another type of force,
i.e., the Lorentz force, in the presence of the magnetic field.
This force is the transverse one to the velocity of the 
particle and, historically, has led to the concept of the 
``field" and relativistic mechanics. 
Then the natural question to 
ask is whether the analogous force exists for light. 
It has been believed that there is no optical analog of 
magnetic field \cite{datta}\cite{born}, namely a Lorentz force on light 
in real space. 
On the other hand, 
a Lorentz force in $k$-space was introduced recently 
in a polarization-dependent equation of motion as a generalization 
of the geometrical optics \cite{onoda}. 
This  ``Lorentz force" in momentum space is described by a Berry phase 
in analogy to an anomalous Hall effect \cite{onoda}
or a Magnus effect \cite{bliokh}. 
Haldane and Raghu focus on an edge state in a magnetic photonic crystal 
on the analogy of the quantum Hall effect \cite{haldane}. 
The edge state is a polarization-dependent one-way waveguide 
with broken time-reversal symmetry. However the Lorentz force
on light in real space has been missing up to now.

In this Letter, we theoretically propose that the 
Lorentz force on light independent of its 
polarization is realized in the multiferroic materials 
showing the non-reciprocal optical effect. 
The polarization-independent non-reciprocal optical effect
requires a breaking of both inversion and time-reversal symmetries,
which allows 
polarization-independent directional birefringence, 
called a magneto-chiral (MC) effect \cite{barron}\cite{rikken} or 
an optical magneto-electric (OME) effect \cite{kubota}\cite{murakami}. 
The difference between the MC and OME effects lies in 
their crystal structures---whether a magnet has a helical structure 
or a polar structure. 
These effects have directional dependence, 
namely, ${\hat k}$ dependence in optical response.
We focus on magnets with a polar structure, namely, multiferroics.  
Novel optical properties in multiferroics 
are characterized by a toroidal moment 
$\vec{T} \equiv \sum_i \vec{r}_i \times \vec{S}_i$ \cite{popov}, 
where $\vec{r}_i$ and $\vec{S}_i$ are
the displacement of the center position from atoms
 and a spin of an electron 
at $i$th site, respectively. 
We note that the definition of the toroidal moment implies that 
even in a antiferroelectric and antiferromagnetic system, 
the toroidal moment can be non zero. 
In ferroelectric and ferromagnetic systems, 
we can rewrite the toroidal moment as 
$\vec{T}\propto \vec{P}\times \vec{M}$, 
where $\vec{P}$ and $\vec{M}$ are spontaneous electric 
polarization and magnetization, respectively.
Hereafter we assume that 
multiferroics are simultaneously ferroelectric and ferromagnetic for 
simplicity. 
We can reverse the toroidal moment by 
an external field by reversing the electric polarization 
or the magnetization.
The toroidal moment brings about the OME effect that 
refractive index depends on whether the propagation direction is 
parallel or antiparallel to the toroidal moment. 
The OME effect is expressed as 
$n_\rightarrow -n_\leftarrow \propto {\hat k}\cdot \vec{T}$, 
where arrows in the subscripts represent directions of light 
and ${\hat k}\equiv \frac{\vec{k}}{|\vec{k}|}$.

We demonstrate below 
a novel refraction phenomenon caused by 
an inhomogeneous toroidal moment in multiferroics. 
The equation of motion for light in multiferroics contains 
a Lorentz force in real space, where 
a spatially inhomogeneous toroidal moment 
gives rise to a magnetic field for light. 
Such spatial modulation of the toroidal moment can be 
achieved in a domain wall in multiferroics. 
From optical viewpoints, 
the toroidal moment plays the role of 
a `vector potential of light. 

Since the OME effect is independent of polarization, 
we can use geometrical optics, which is a scalar representation, 
to describe the wave propagation through multiferroics.
The OME effect is described by 
\be
n(\vec{r}) =n_0(\vec{r}) + \alpha {\hat k}\cdot \vec{T}(\vec{r}), 
\label{n}
\en
where ${\hat k}={\vec k}/|{\vec k}|$ is now 
represented as ${\hat k}=\frac{d\vec{r}}{ds}$. 
Substituting the refractive 
index (\ref{n}) into  Eq.~(1), we obtain 
the following non-linear equation, 
\bega
\frac{d}{ds} \bigl[ n_0(\vec{r}){\hat k} \bigr] 
+\alpha \frac{d}{ds}\bigl[ ({\hat k}\cdot 
\vec{T}(\vec{r})){\hat k} \bigr]
={\rm grad}\ n_0(\vec{r})\nonumber \\
+\alpha({\hat k}\cdot \nabla )\vec{T}(\vec{r})
+\alpha {\hat k} \times {\rm rot}\vec{T}(\vec{r}).
\label{nonlinear} 
\eda
Each term on the 
Eq.~(\ref{nonlinear}) can be interpreted as follows. 
The first term on the right-hand side represents 
a ``potential force" for photons which yields 
a conventional Snell's law. 
The second term is the differential of 
the toroidal moment along the 
propagation direction, whose physical meaning is explained later. 
Remarkably, 
the third term is nothing but a Lorentz force 
on light with a vector potential $\vec{T}(\vec{r})$. 
The ${\hat k}$ dependence in the second and the third terms 
gives rise to non-reciprocal refraction, namely, one-way propagation. 

\begin{figure}[h]
  \centerline{
    \epsfxsize=8.5cm
    \epsfbox{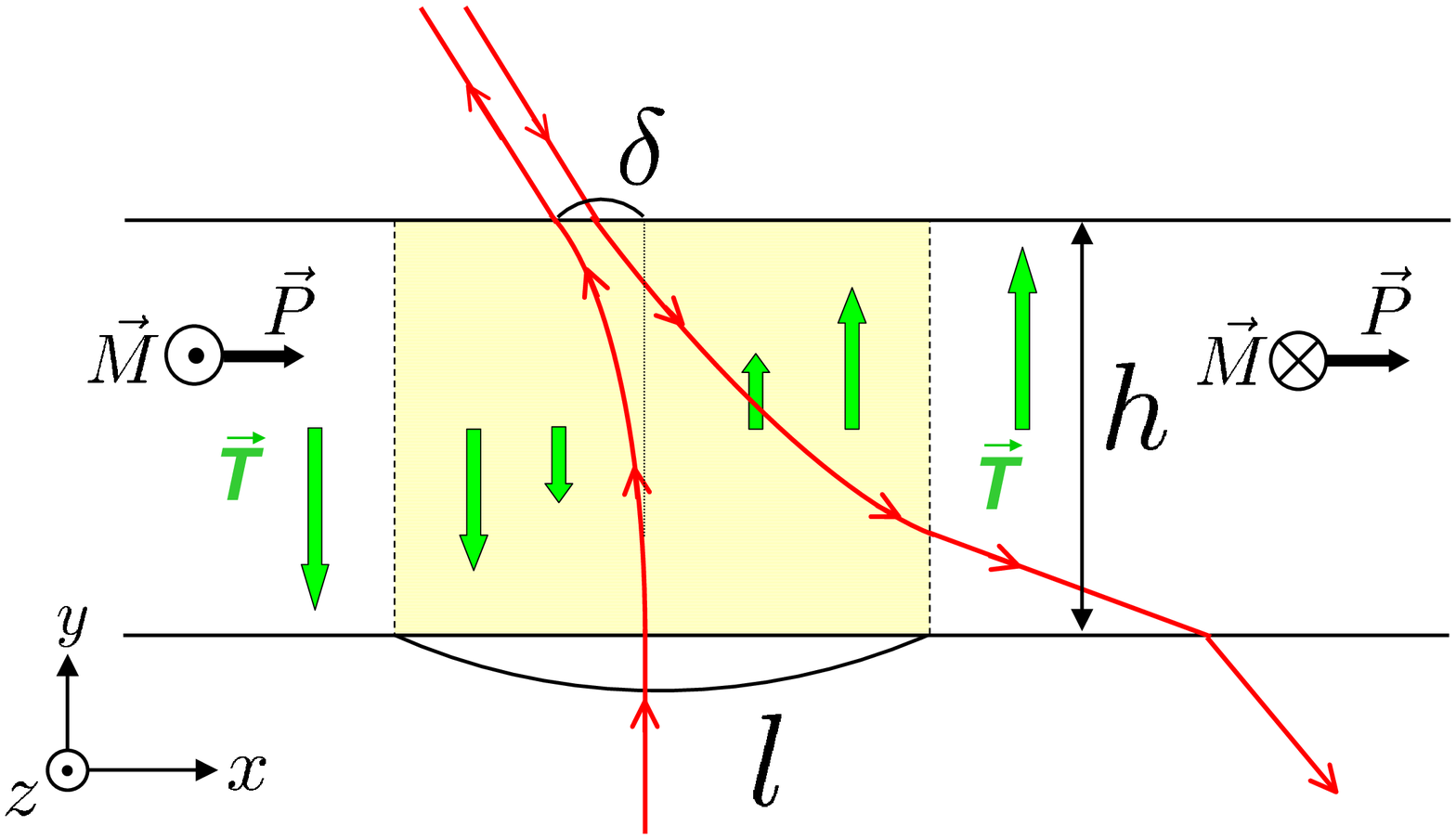}
  }
  \caption{Non-reciprocal refraction through a toroidal domain wall 
represented by a yellow box. The green 
arrows represent toroidal moments. 
Different propagation direction has different optical path, 
represented by the lines with red arrows.}
 \label{refraction}
\end{figure}

In realistic materials, 
the contribution of the Lorentz force term 
can be observed in a domain wall(DW) of the toroidal moment.
In a toroidal domain-wall(TDW) region, the toroidal moment varies as 
$\vec{T}=(0, Bx, 0)$, which is analogous to the 
vector potential of a uniform magnetic field $B$ in the Landau 
gauge, as shown in Fig.~\ref{refraction}.
Since the toroidal moment is roughly related to the electric polarization 
and the magnetization as $\vec{T}\propto \vec{P}\times \vec{M}$, 
the toroidal moment can be reversed by the 
reversal of $\vec{P}$ or $\vec{M}$ by an external field. 
In the TDW, we have $n_0(\vec{r})=n_0$ and
the equation is explicitly written as 
\begin{eqnarray*}
n_0\ddot{x}+\alpha B\frac{d}{ds}(\dot{y}x\dot{x})
-\alpha B\dot{y}=0, \\
n_0\ddot{y}+\alpha B\frac{d}{ds}(x\dot{y}^2)=0, 
\end{eqnarray*}
where the dot denotes a derivative with respect to $s$.
We assume the initial conditions to be 
$\vec{r} \Bigl|_{s=0} =(0, 0, 0)$, 
${\hat k} \Bigl|_{s=0} =(\cos \theta, \sin \theta, 0)$, 
and we can integrate the equations; 
\begin{eqnarray*}
n_0\dot{x}+\alpha B\dot{y}x\dot{x}-\alpha B y =n_0 \cos \theta, \\
n_0\dot{y}+\alpha Bx\dot{y}^2 =n_0 \sin \theta. 
\end{eqnarray*}
Then the solution to these equations is  
\be
x=\frac{B\alpha}{2n_0 \sin \theta} y^2 
+\frac{\cos \theta}{\sin \theta}y,
\en 
which is interpreted as follows. 
The Lorentz force term of the right-hand side on Eq.~(\ref{nonlinear}) 
is explicitly written to be
 ${\hat k} \times {\rm rot}\ \vec{T}=
({\hat k}_yB, -{\hat k}_xB, 0)
$, which gives rise to a cyclotron motion. 
The second term gives 
$({\hat k} \cdot \nabla ) \vec{T}=(0, +{\hat k}_xB, 0)$ 
which cancels the $y$ component of the Lorentz term. 
Therefore the right-hand side in Eq.~(\ref{nonlinear}) is
proportional to $({\hat k}_yB, 0, 0)$, 
leading to the parabolic solution.
When the direction of the initial velocity is reversed, 
a sign of a curvature of the parabola is changed, 
which means that we have a one-way trajectory through a TDW. 
Because the toroidal moment is a physical field 
for optical waves and is related to 
the electric polarization and the magnetization, 
it has no gauge invariance and can be controlled by 
external fields or temperature. 
It means that one can design a vector potential 
at one's disposal by spatially modulating $\vec{P}$ or $\vec{M}$. 

Let us estimate the displacement $\delta$ in Fig.~1. 
We put the thickness and the height of a sample to be 
$l$ and $h$, respectively. 
The displacement is roughly estimated as 
$\delta \sim (h^2/l)a$, 
where $a$ is the strength of the OME effect; 
$a=(n_\rightarrow -n_\leftarrow)/n_0$. 
In GaFeO$_3$, these parameters are 
$l \sim 100$ nm, $h \sim 10$ $\mu$m and $a\sim 10^{-3}$. 
Using these parameters, we obtain the displacement to be 
$\delta \sim 1\mu$m\cite{note}, 
which can be observed through a microscope. 
A condition for large $\delta$ is to have a large amount of 
magnetic field and a large medium. 
Such a strong magnetic field can be achieved in a material 
with a large OME effect and a thin TDW, namely large $a$ 
and small $l$, respectively. 
A large OME effect gives rise to a large refraction, 
and small $l$ corresponds to a large 
differential of the toroidal moment, resulting in a large displacement.
Here the validity of the geometrical optics should be discussed.
Since the thickness of the DW in GaFeO$_3$ is 100nm 
which is shorter than 
the wavelength of the visible light, 
the refraction of the light would be obscure. 
Therefore we should use ultraviolet beam. 
However, 
a material with a DW whose thickness is larger than the 
wavelength of the incident light might be a good candidate 
to observe this non-reciprocal refraction.

\begin{figure}[h]
  \centerline{
    \epsfxsize=8.5cm
    \epsfbox{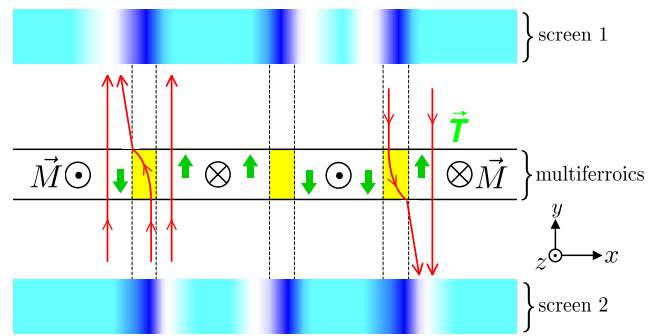}
  }
  \caption{Non-reciprocal refraction through some domain walls. 
Domain walls are observed as contrasts on a screen, represented by 
deep blue, bright blue, and white regions. The two screens 1 and 2 
with different contrasts correspond to the incident directions, 
the face and the back, respectively.}
\label{contrasts}
\end{figure}

As an example, 
we consider an experimental set up for 
the TDWs shown in Fig.~\ref{contrasts}. 
Here, we introduce the TDW as a magnetic DW with 
uniform electric polarization. 
The incident optical ray is refracted through each DW, 
leading to two distinct regions as contrasts in a screen.
One is a dark region represented by dark blue 
where photons cannot reach. 
The other is a bright region characterized by a white color 
where both refracted light and transmitted light can reach. 
It means that TDWs can be observed as contrasts in a screen \cite{note2}.
The region in which the light propagates out of TDWs is represented by 
a bright blue color.
The contrasts purely have their origins 
in the spatially varying toroidal moment. 
To examine whether the contrasts are really 
caused by the inhomogeneous toroidal moment 
or not, we propose following two experimental methods. 
One is to introduce the light from another side of the sample. 
There should be different contrasts from those of another side, 
represented by two screens 1 and 2 in Fig.~\ref{contrasts}.
The other is to apply the external magnetic field $\sim 500$ Oe 
parallel to $z$ direction to order the magnetization. 
The magnetic DW and the TDW vanish, 
resulting in the disappearance of the contrasts. 
Both of these two experimental tests 
are needed to confirm our proposal.

The OME effect is usually very small for direct observation; 
$n_\rightarrow -n_\leftarrow \sim$
$10^{-3}$ for GaFeO$_3$ \cite{kubota}, 
$\sim 10^{-6}$ for Cr$_2$O$_3$ \cite{krich}, 
$\sim 10^{-4}$ for Er$_{1.5}$Y$_{1.5}$Al$_5$O$_{12}$ \cite{rikken2} and so on. 
To overcome the disadvantage, 
one can enhance the OME effect in artificial structures. 
In periodic structures composed of multiferroics, 
the OME effect can be magnified by a factor of 1000 
in multiferroic gratings and photonic crystals \cite{sawada}. 
In contrast, what we demonstrate here provides another example for 
a sizable OME effect. 
The non-reciprocal refraction obtained in this Letter 
is of the order of 1 $\mu$m, which can be observed 
without any use of a cavity, a photonic crystal or a patterned structure.
Therefore, non-reciprocal refraction due to 
a toroidal moment modulation provides us with 
a new method to measure the OME effect.

Kida \textit{et. al.} found an OME effect in a submicron 
patterned magnet in which inversion symmetry is artificially 
broken by a chevron-shaped structure \cite{kida}. 
The signal of the OME effect in the patterned magnet 
is of the order of $10^{-3}$, which is significantly large. 
They show that we can introduce 
a toroidal moment in a magnet with an artificially asymmetric 
structure. 
It may be possible that 
an inhomogeneous toroidal moment introduced by an artificial structure 
yields a Lorentz force on light. 

In conclusion, we have investigated a novel non-reciprocal refraction 
in multiferroics, which is polarization independent.
The toroidal moment, inherent in multiferroics, 
plays the role of vector potential for light, and
its spatial variation gives optical analogue of a Lorentz force. 
Photons feel no Lorentz force in a conventional sense 
under a real external magnetic field
because they have no electric charge. 
The toroidal moment is a physical field without gauge degrees of freedom, 
and can be experimentally controlled by an external field or 
temperature. 
It means that the Lorentz force can also be 
controlled by them. 
Such an interesting material as a multiferroic matter 
has the potential importance of a novel external field for photons; 
the Lorentz force obtained here is one example.
This Lorentz force on light fills one of the most 
fundamental pieces which has been missing for 
photon-electron analogy.

The authors are grateful to 
T. Arima, M. Onoda, N. Kida, S. Murakami, H. Katsura,
 and Y. Tokura for fruitful discussions. 
K.S. is supported by a Japan Society for the Promotion of Science. 
This work is financially supported by a NAREGI Grant, Grant-in-Aids 
from the Ministry of Education, Culture, Sports, Science and Technology 
of Japan.



\begin{thebibliography}{99}

\bibitem{datta} S. Datta, 
\textit{Electronic Transport in Mesoscopic Systems}, 
(Cambridge University Press, 1995).

\bibitem{born}M. Born and E. Wolf, 
\textit{Principles of Optics}, 7th  edition 
(Cambridge University Press, 1999).

\bibitem{onoda} M. Onoda, S. Murakami and N. Nagaosa, \phrl{93}{083901}{2004}.

\bibitem{bliokh}K. Yu. Bliokh and Yu. P. Bliokh, \phre{70}{026605}{2004}; 
\jetpl{79}{519}{2004}.

\bibitem{haldane}F. D. M. Haldane and S. Raghu, cond-mat/0503588.

\bibitem{barron}L. D. Barron and J. Vrbancich, \mol{51}{715}{1984}.

\bibitem{rikken} G. L. J. A. Rikken and E. Raupach, 
\nature{390}{493}{1997}; 
G. L. J. A. Rikken and E. Raupach, Phys. Rev. E \textbf{58}, 5081 (1998); 
C. Koerdt, G. D\"uchs and G. L. J. A. Rikken, 
Phys. Rev. Lett. \textbf{91}. 073902 (2003); 
F. A. Pinheiro and B. A. van Tiggelen, 
J. Opt. Soc. Am. A \textbf{20}, 99(2003).

\bibitem{kubota} M. Kubota, T. Arima, Y. Kaneko, 
J. P. He, X. Z. Yu and Y. Tokura, 
Phys. Rev. Lett. \textbf{92}. 137401 (2004); 
J. H. Jung, M. Matsubara, T. Arima, J. P. He, Y. Kaneko 
and Y. Tokura, 
Phys. Rev. Lett. \textbf{93} 037403 (2004); 
T. Arima, J. H. Jung, M. Matsubara, M. Kubota, J. P. He, 
Y. Kaneko and Y. Tokura, \jpsj{74}{1419}{2005}.

\bibitem{murakami} S. Murakami, R. Shindou, N. Nagaosa and A. S. Mishchenko, 
\phrb{66}{184405}{2002}.

\bibitem{popov} Yu. F. Popov, A. M. Kadomtseva, G. P. Vorob'ev, V. A. Timofeeva, D. M. Ustinin, A. K. Zvezdin and M. M. Tegeranchi, \jetp{87}{146}{1998}. 

\bibitem{note} For GaFeO$_3$, optical path in Fig.~1 is not the case. 
The thickness $l$ is so small in GaFeO$_3$ that optical beam 
is refracted and out of a TDW. The trajectory of 
the beam is a parabola in a TDW and becomes a line out of a TDW. 
The naive estimation $\delta \sim \frac{h^2}{l}a$ breaks down. 
However, careful calculation suggests that 
the displacement still holds $\delta \sim 1\mu$m.
 
\bibitem{note2} These contrasts simply represent whether 
photons reach or not, regardless of interference due to 
difference in light paths of refracted one and unrefracted one. 
Coherent light may make the interference pattern on a screen 
in addition to the contrasts. 

\bibitem{krich}B. B. Krichevtsov, V. V. Pavlov, R. V. Pisarev, and V. N. Gridnev, \jpcm{5}{8233}{1993}.

\bibitem{rikken2} G. L. J. A. Rikken, C. Strohm and P. Wyder, 
\phrl{89}{133005}{2002}.

\bibitem{sawada} K. Sawada and N. Nagaosa, \appl{87}{042503}{2005}.

\bibitem{kida} N. Kida, T. Yamada, M. Konoto, Y. Okimoto, T. Arima, K. Koike, H. Akoh and Y. Tokura, \phrl{94}{077205}{2005}.
\end{thebibliography}
\end{document}